\documentclass[conference]{IEEEtran}
\IEEEoverridecommandlockouts

\usepackage{graphicx,cite,acronym,amsmath,amssymb,amsfonts,color,floatflt,stfloats}
\usepackage{caption}
\usepackage{subcaption}
\usepackage{epsfig,epstopdf,psfrag}
\usepackage[ruled,vlined]{algorithm2e}
\SetKwInput{KwInput}{Input}                
\SetKwInput{KwOutput}{Output}  

\usepackage[table]{xcolor}
\usepackage{bm}

\usepackage[cmintegrals]{newtxmath}
\usepackage{graphics,hhline,array,cite,bbm}

\usepackage{latexsym}
\usepackage{theorem}
\usepackage{times}
\usepackage{makeidx}

\DeclareGraphicsExtensions{.png,.jpg,.eps}

\theoremstyle{plain}

\newtheorem{prop}{Proposition}

\DeclareMathOperator{\tr}{tr}

\usepackage{bbm}

\newcommand{\mbf}{\mathbf}

\IEEEoverridecommandlockouts

\usepackage{tikz}

\usetikzlibrary{positioning} 
\definecolor{Eored}{rgb}{.647 ,.129 ,.149} 
\definecolor{Eogreen}{rgb}{0 ,0.53 ,0}

\begin{document}

\title{\Large MIMO Interference Channels Assisted by Reconfigurable Intelligent Surfaces: Mutual Coupling Aware Sum-Rate Optimization Based on a Mutual Impedance Channel Model \vspace{-0.50cm}}
\author{Andrea~Abrardo, \IEEEmembership{Senior~Member,~IEEE}, Davide~Dardari, \IEEEmembership{Senior~Member,~IEEE}, \\ Marco~Di~Renzo, \IEEEmembership{Fellow,~IEEE}, and Xuewen~Qian
\vspace{-1.0cm}
\thanks{Manuscript received Feb. 14, 2021. A. Abrardo is with the Univerity of Siena and CNIT, Italy (e-mail: abrardo@dii.unisi.it). D. Dardari is with the University of Bologna and CNIT, Italy (e-mail: davide.dardari@unibo.it). M. Di Renzo and X. Qian are with CNRS and Paris-Saclay University, France (e-mail: marco.di-renzo@universite-paris-saclay.fr).}
}


\markboth{TBA} {A. Abrardo, D. Dardari, M. Di Renzo, and X. Qian, MIMO Interference Channels Assisted by Reconfigurable Intelligent Surfaces: Mutual Coupling Aware Sum-Rate Optimization Based on a Mutual Impedance Channel Model }

\maketitle

\begin{abstract}
We investigate a multi-user multiple-input multiple-output interference network in the presence of multiple reconfigurable intelligent surfaces (RISs). The entire system is described by using a circuit-based model for the transmitters, receivers, and RISs. This is obtained by leveraging the electromagnetic tool of mutual impedances, which accounts for the signal propagation and the mutual coupling among closely-spaced scattering elements. An iterative and provably convergent optimization algorithm that maximizes the sum-rate of RIS-assisted multi-user interference channels is introduced. Numerical results show that the sum-rate is enhanced if the mutual coupling among the elements of the RISs is accounted for at the optimization stage.
\end{abstract}
\begin{IEEEkeywords}
Reconfigurable intelligent surfaces, mutual impedances, mutual coupling, sum-rate, optimization.
\end{IEEEkeywords}

\section{Introduction}
\label{Sec:Introduction}
A reconfigurable intelligent surface (RIS) is a nearly-passive thin sheet of electromagnetic material that can make a complex radio environment programmable at the electromagnetic level \cite{MDR_JSAC}. To evaluate the performance benefits and to optimize the deployment and operation of RIS-assisted wireless networks, it is necessary to utilize channel and communication models that account for the electromagnetic characteristics and the physical implementation of the RISs. This is an open research issue that is currently subject to intense investigation \cite{Gabriele_HE2E}.

The authors of \cite{Gabriele_HE2E}, in particular, have recently introduced an electromagnetic-compliant communication model for RIS-assisted communications, which resembles a multiple-input multiple-output (MIMO) channel. The model proposed in \cite{Gabriele_HE2E} can be applied to an RIS made of closely-spaced scattering elements that are controlled via tunable impedances. The electromagnetic field scattered by the RIS is engineered through an appropriate design of the tunable impedances. By departing from the channel model in \cite{Gabriele_HE2E}, the authors of \cite{Xuewen_HE2E} have recently introduced an analytical framework and a numerical algorithm that optimize the tunable impedances so as to maximize the received power. It is shown that major gains are obtained if the electromagnetic properties (e.g., the mutual coupling) and the circuital implementation (e.g., the tunable impedances) of the RIS are taken into account at the optimization stage.

The algorithm introduced in \cite{Xuewen_HE2E} is, however, applicable only to single-antenna transmitters and receivers. In addition, a single RIS and a single receiver are considered. In this paper, we depart from the channel model introduced in \cite{Gabriele_HE2E} and introduce an algorithm for optimizing an RIS-assisted wireless network in the presence of an arbitrary number of multi-antenna transmitters, multi-antenna receivers, and RISs that are shared among all the transmitter-receiver pairs. Notably, the interference among all the available transmitter-receiver pairs is taken into account (MIMO interference channel). The proposed algorithmic solution leverages the weighted minimum mean square error (wMMSE) algorithm and the iterative block coordinate descent (BCD) method. The obtained results show that the sum-rate is enhanced if the mutual coupling among the elements of the RISs is accounted for at the design stage.

\textit{Notation}: $\Re$ and $\Im$ are real and imaginary parts; $\mathbb{E}$ is the expectation; $(\cdot)^H$, $(\cdot)^T$, $\tr(\cdot)$ are Hermitian, transpose, trace; $\odot$ is the Hadamard product; $\left\| \cdot \right \|$ is the spectral norm; $\nabla$ is the gradient; $\mathbf{I}_{N}$, $\mathbf{0}_{N}$ are the $N \times N$ identity and all-zero matrices.

\section{System and Signal Model}
We consider a MIMO interference channel that comprises $N_u$ transmitter-receiver pairs. Each transmitter is equipped with $M$ antennas and each receiver is equipped with $L \le M$ antennas. Based on \cite{Gabriele_HE2E}, each antenna element of the transmitter and receiver is assumed to be a thin wire dipole of perfectly conducting material. Each transmit antenna element is driven by an independent voltage generator that models a transmit feed line, and each receive antenna element is connected to a load impedance that models a receive electric circuit. 

For simplicity, we assume that the number of symbols (independent streams) sent by each transmitter is equal to the number of receive antennas. The transmission between the $N_u$ transmitter-receiver pairs is assisted by $K$ RISs. Each RIS comprises $P$ nearly-passive tiny scattering elements that can be independently configured through a network controller. We use the indices $j$, $k$, and $i$ to denote the $j$th transmitter, $k$th RIS, and $i$th receiver. With this notation, we imply that the intended receiver of the $j$th transmitter is the $i$th receiver. The $K$ RISs are shared among all the $N_u$ transmitter-receiver pairs.

We denote by $\mathbf{s}_j = \left[s_{j}(1),s_{j}(2),\ldots,s_{j}(L)\right]^T$ the complex vector that comprises the $L$ information symbols of the $j$th transmitter. The information symbols are assumed to be zero-mean and independent and identically distributed (i.i.d.) random variables (RVs), i.e., $\mathbb{E}\left[\mathbf{s}_j \mathbf{s}_j^H\right] =  \mathbf{I}_{L}$ and $\mathbb{E}\left[\mathbf{s}_j \mathbf{s}_i^H\right] =  \mathbf{0}_{L}$ for $j \ne i$. Denoting by $\mathbf{V}_j \in \mathbb{C}^{M \times L}$ the precoding matrix of the $j$th transmitter, its transmitted vector is $\mathbf{x}_j = \mathbf{V}_j \mathbf{s}_j \in \mathbb{C}^{M \times 1}$.

As for the RISs, we adopt the electromagnetic-compliant communication model recently introduced in \cite{Gabriele_HE2E}, which is based on mutual impedances. The channel model in \cite{Gabriele_HE2E} is applicable to RISs constituted by an array of thin wire dipoles of perfectly conducting material. Each dipole is controlled by a tunable load impedance that enables the control of the scattered field. Thus, the RIS-assisted channel can be appropriately programmed and shaped by optimizing the tunable impedances. In \cite[Theorem 1]{Gabriele_HE2E}, the authors introduce an $L \times M$ end-to-end channel matrix that formulates the voltage measured at the ports of the receive antennas as a function of the voltage generators connected to the ports of the transmit antennas. The channel matrix is applicable to a general communication system, which encompasses a multi-antenna transmitter, a multi-antenna receiver, and the possibility that all the radiating elements are in the near-field of each other. In this paper, we consider a simplified case study in which the $M$-antenna transmitters and the $L$-antenna receivers are in the far-field of each other and in the far-field of the RISs. However, the $P$ tiny scattering elements that comprise each RIS can be arbitrarily close to each other, and the mutual coupling among them is appropriately taken into account. 

Based on \cite[Theorem 1]{Gabriele_HE2E}, the $L \times M$ channel matrix (denoted by ${\bf{H}}_{i,j}^{\left( k \right)}$) between the $j$th transmitter and the $i$th receiver, which accounts for the line-of-sight link between them and the scattered link from the $k$th RIS, can be formulated as
\vspace{-0.2cm}
\begin{align}
{\bf{H}}_{i,j}^{\left( k \right)} =& {\left( {{{\bf{I}}_{L}} + {\bf{\Psi }}_{i,i}^{\left( k \right)}{\bf{Z}}_i^{ - 1} - {\bf{\Psi }}_{i,j}^{\left( k \right)}{{\left( {{\bf{\Psi }}_{j,j}^{\left( k \right)} + {{\bf{Z}}_j}} \right)}^{ - 1}}{\bf{\Psi }}_{j,i}^{\left( k \right)}{\bf{Z}}_i^{ - 1}} \right)^{ - 1}} \nonumber \\
& \times {\bf{\Psi }}_{i,j}^{\left( k \right)}{\left( {{\bf{\Psi }}_{j,j}^{\left( k \right)} + {{\bf{Z}}_j}} \right)^{ - 1}} \quad \in \mathbb{C}^{L \times M} \label{HE2E}
\end{align}
\noindent where ${{{\bf{Z}}_j}}$ and ${{{\bf{Z}}_i}}$ are $M \times M$ and $L \times L$ diagonal matrices that collect the internal impedances of the transmit generators and the load impedances of the receive antennas, respectively, and 
\vspace{-0.4cm}
\begin{align}
 & \hspace{-0.75cm} {\bf{\Psi }}_{j,j}^{\left( k \right)} = {{\bf{Z}}_{j,j}} - {{\bf{Z}}_{j,k}}{\left( {{{\bf{Z}}_{k,k}} + {{\bf{Z}}_{{\rm{tun}}}^{(k)}}} \right)^{ - 1}}{{\bf{Z}}_{k,j}} \quad \in \mathbb{C}^{M \times M}  \\
& \hspace{-0.75cm} {\bf{\Psi }}_{j,i}^{\left( k \right)} = {{\bf{Z}}_{j,i}} - {{\bf{Z}}_{j,k}}{\left( {{{\bf{Z}}_{k,k}} + {{\bf{Z}}_{{\rm{tun}}}^{(k)}}} \right)^{ - 1}}{{\bf{Z}}_{k,i}} \quad \in \mathbb{C}^{M \times L}  \\
& \hspace{-0.75cm} {\bf{\Psi }}_{i,j}^{\left( k \right)} = {{\bf{Z}}_{i,j}} - {{\bf{Z}}_{i,k}}{\left( {{{\bf{Z}}_{k,k}} + {{\bf{Z}}_{{\rm{tun}}}^{(k)}}} \right)^{ - 1}}{{\bf{Z}}_{k,j}} \quad \in \mathbb{C}^{L \times M}  \\
& \hspace{-0.75cm} {\bf{\Psi }}_{i,i}^{\left( k \right)} = {{\bf{Z}}_{i,i}} - {{\bf{Z}}_{i,k}}{\left( {{{\bf{Z}}_{k,k}} + {{\bf{Z}}_{{\rm{tun}}}^{(k)}}} \right)^{ - 1}}{{\bf{Z}}_{k,i}} \quad \in \mathbb{C}^{L \times L} 
\end{align}
\noindent where ${{\bf{Z}}_{x,y}}$, for $x,y \in \left\{ {j,k,i} \right\}$, is the matrix of mutual (or self if $x=y$) impedances between the radiating elements of $y$ and $x$, which characterizes the signal propagation and the mutual coupling between $x$ and $y$, and ${{\bf{Z}}_{{\rm{tun}}}^{(k)}}$ is the $P \times P$ diagonal matrix of tunable impedances of the $k$th RIS. The matrices ${{\bf{Z}}_{x,y}}$ can be computed by using \cite[Lemma 2]{Gabriele_HE2E}, which shows that they depend only on the geometry and the physical implementation of the RISs, e.g., the scattering elements of the RISs are thin wire dipoles. In this paper, therefore, the impedances ${{\bf{Z}}_{x,y}}$ need to be computed only once and are assumed to be fixed and given parameters, while the matrix ${{\bf{Z}}_{{\rm{tun}}}^{(k)}}$, which ensures the reconfigurability of the $k$th RIS, is a variable that is optimized to maximize the system sum-rate.

If the transmitters, the receivers, and the RISs are in the far-field of each other, while still taking the mutual coupling among the closely-spaced scattering elements of each RIS into account, \eqref{HE2E} can be simplified. The self impedances ${{\bf{Z}}_{x,x}}$ are, in fact, independent of the transmission distances of the transmitter-receiver, transmitter-RIS, and RIS-receiver links, and they depend only on the inter-distances between the radiating elements that comprise each transmitter, RIS, and receiver. In the far-field region, therefore, we have ${\bf{\Psi }}_{j,j}^{\left( k \right)} \approx {{\bf{Z}}_{j,j}}$, ${\bf{\Psi }}_{i,i}^{\left( k \right)} \approx {{\bf{Z}}_{i,i}}$, and ${\bf{\Psi }}_{i,i}^{\left( k \right)}{\bf{Z}}_i^{ - 1} - {\bf{\Psi }}_{i,j}^{\left( k \right)}{\left( {{\bf{\Psi }}_{j,j}^{\left( k \right)} + {{\bf{Z}}_j}} \right)^{ - 1}}{\bf{\Psi }}_{j,i}^{\left( k \right)}{\bf{Z}}_i^{ - 1} \approx {\bf{\Psi }}_{i,i}^{\left( k \right)}{\bf{Z}}_i^{ - 1}$. If the radiating elements of each transmitter and each receiver are sufficiently spaced apart, the matrices ${{\bf{Z}}_{j,j}}$ and ${{\bf{Z}}_{i,i}}$ are almost diagonal matrices, i.e., the off-diagonal entries are much smaller than the diagonal entries. Due to the small inter-distances between the scattering elements of the RISs, on the other hand, the matrices ${{{\bf{Z}}_{k,k}}}$ are, in general, full matrices. In the far-field region, thus, $\mathbf{H}_{i,j}^{(k)}$ in \eqref{HE2E} can be approximated as
\vspace{-0.2cm}
\begin{align}
\hspace{0.25cm} {\bf{H}}_{i,j}^{\left( k \right)} &\approx {\left( {{{\bf{I}}_{L}} + {{\bf{Z}}_{i,i}}{\bf{Z}}_i^{ - 1}} \right)^{ - 1}}{{\bf{Z}}_{i,j}}{\left( {{{\bf{Z}}_{j,j}} + {{\bf{Z}}_j}} \right)^{ - 1}} \\
 & \hspace{-0.5cm}- {\left( {{{\bf{I}}_{L}} + {{\bf{Z}}_{i,i}}{\bf{Z}}_i^{ - 1}} \right)^{ - 1}}{{\bf{Z}}_{i,k}}{\left( {{{\bf{Z}}_{k,k}} + {\bf{Z}}_{{\rm{tun}}}^{\left( k \right)}} \right)^{ - 1}}{{\bf{Z}}_{k,j}}{\left( {{{\bf{Z}}_{j,j}} + {{\bf{Z}}_j}} \right)^{ - 1}} \nonumber \, .
\end{align}

By considering the contribution of the $K$ RISs available in the system and by introducing the shorthand notation
\vspace{-0.2cm}
\begin{align}
& {{{\bf{\bar H}}}_{i,j}} = {\left( {{{\bf{I}}_{L }} + {{\bf{Z}}_{i,i}}{\bf{Z}}_i^{ - 1}} \right)^{ - 1}}{{\bf{Z}}_{i,j}}{\left( {{{\bf{Z}}_{j,j}} + {{\bf{Z}}_j}} \right)^{ - 1}} \quad \in \mathbb{C}^{L \times M} \\
& {{\bf{T}}_{i,k}} = {\left( {{{\bf{I}}_{L}} + {{\bf{Z}}_{i,i}}{\bf{Z}}_i^{ - 1}} \right)^{ - 1}}{{\bf{Z}}_{i,k}} \quad \in \mathbb{C}^{L \times P}\\
& {{\bf{S}}_{k,j}} = {{\bf{Z}}_{k,j}}{\left( {{{\bf{Z}}_{j,j}} + {{\bf{Z}}_j}} \right)^{ - 1}} \quad \in \mathbb{C}^{P \times M}\\
& {{{\bf{\bar B}}}_k} = {{\bf{Z}}_{k,k}} \quad \in \mathbb{C}^{P \times P}, \quad {{\bf{B}}_k} = {\bf{Z}}_{{\rm{tun}}}^{\left( k \right)} \quad \in \mathbb{C}^{P \times P}\\
& {{{\bf{\tilde H}}}_{i,k,j}} =  - {{\bf{T}}_{i,k}}{\left( {{{{\bf{\bar B}}}_k} + {{\bf{B}}_k}} \right)^{ - 1}}{{\bf{S}}_{k,j}} \quad \in \mathbb{C}^{L \times M}\label{HB}
\end{align}
\noindent the end-to-end channel matrix from the $j$th transmitter to the $i$the receiver can be formulated, in the far-field region, as
\begin{equation}
\begin{array}{c}
\mathbf{H}_{i,j}\left(\mathcal{B}\right) \approx \mathbf{{\bar{H}}}_{i,j} + \sum\nolimits_{k=1}^{K} \mathbf{\tilde{H}}_{i,k,j}\left(\mathcal{B}\right) 
\quad \in \mathbb{C}^{L \times M}
\end{array}
\label{eq:rx_irs3}
\end{equation}
\noindent where ${{{\bf{\bar H}}}_{i,j}}$ accounts for the line-of-sight link and ${{{\bf{\tilde H}}}_{i,k,j}}$ accounts for the (virtual line-of-sight) link scattered by the $k$th RIS. In \eqref{eq:rx_irs3}, we have made explicit the dependence of the scattered field with the diagonal matrix $\mathbf{B}_{k} = \text{diag}\left(\mathbf{b}_{k}\right)$ of tunable impedances of the $K$ RISs. In particular, $\mathcal{B}=\left\{\mbf{b}_{1},\mbf{b}_{2},\dots,\mbf{b}_{K}\right\}$ denotes the set of $K$ vectors $\mathbf{b}_{k}$ to be optimized. For simplicity, only single reflections from the RISs are considered in \eqref{eq:rx_irs3}.

By taking into account the concurrent transmissions of the $N_u$ transmitters, the signal at the $i$th receiver is
\begin{equation}
\begin{array}{c}
\mathbf{y}_{i} = \mathbf{H}_{i,i}\left(\mathcal{B}\right) {\bf{x}}_i  + \sum\nolimits_{j=1, j \neq i}^{N_u } \mathbf{H}_{i,j}\left(\mathcal{B}\right)  {\bf{x}}_j  + \mathbf{n}_i 
\quad \in \mathbb{C}^{L \times 1}
\end{array}
\label{eq:rx_total}
\end{equation}
where $\mathbf{n}_i \in \mathbb{C}^{L \times 1}$ denotes the additive white Gaussian noise with distribution $\mathcal{CN}\left(0,\sigma_i^2 \mathbf{I}_L\right)$. 
Based on the resulting MIMO interference channel in \eqref{eq:rx_total}, the achievable rate of the $i$th transmitter-receiver pair can be formulated as \cite{Andrea_wMMSE} 
\begin{equation}
\label{eq:rate2.1}
R_{i}(\mathcal{V},\mathcal{B})= \log \det
\left(
{{\bf{I}}_{L}} +
{{\bf{V}}_i^H{\bf{H}}_{i,i}^H\left(\mathcal{B}\right)\bar{\mbf{J}}_{i}^{ - 1}{\bf{H}}_{i,i}\left(\mathcal{B}\right)}{{\bf{V}}_i}
\right)
\end{equation}
where $\bar{\mbf{J}}_{i}=\sum\nolimits_{j =1,j \neq i}^{N_u} {{{\bf{H}}_{i,j}\left(\mathcal{B}\right)}{{\bf{V}}_j}{\bf{V}}_j^H{\bf{H}}_{i,j}^H\left(\mathcal{B}\right)}  + {\sigma_i ^2}{{\bf{I}}_{L}}$ is the interference-plus-noise covariance matrix  and $\mathcal{V}=\left\{\mbf{V}_{1},\mbf{V}_{2},\dots,\mbf{V}_{N_{u}}\right\}$ denotes the set of $N_{u}$ precoding matrices.

\begin{algorithm}[!t]
\footnotesize
\caption{BCD for RIS optimization}
\textbf{Initialize:} RIS impedances $\mathbf{B}^{(0)}_{k}$; precoding matrices ${\mathbf{V}}_i^{(0)}$;\\ \hspace{1.15cm} small increment $0 \le \delta \ll 1$; number of iterations $\mathcal{Q}$; \

\For {$q = 1,\ldots,\mathcal{Q}$}
{

		Compute ${\mathbf{G}}_{i}^{(q)}$,  ${\mathbf{W}}_{i}^{(q)}$ and ${\mathbf{V}}_{i}^{(q)}$ from Algorithm \ref{Alg:wMMSE};\\ \label{algLine:gwvUpdate}
		
		\For {$k = 1,\ldots,K$}
	        {
		Compute ${\mathbf{{M}}}_{k}$ and ${\mathbf{{u}}}_{k}$ according to \eqref{eq:mMSEMat4.4};\\ \label{algLine:mvUpdate}
		Compute $\boldsymbol{\delta}_{k}$ according to \eqref{eq:precoderShir}\;
		$\mathbf{B}^{(q+1)}_{k}  \leftarrow  \mathbf{B}^{(q)}_{k} + \mathbf{\Delta}_k$\;
		}	
}
\label{Alg:DIA-PD} 
\end{algorithm} \setlength{\textfloatsep}{-60pt}%

\vspace{-0.10cm}
\section{Problem Formulation and Solution} \vspace{-0.15cm}
In this paper, we are interested in optimizing the two sets $\mathcal{V}$ and $\mathcal{B}$ so as to maximize the system sum-rate. Let $P_i$ be the power budget of the $i$th transmitter and $\boldsymbol{\alpha}=[\alpha_{1},\alpha_{2},\dots,\alpha_{N_{u}}]$ be a set of weights that is chosen for ensuring some fairness among the $N_u$ transmitter-receiver pairs \cite{Andrea_wMMSE}. Thus, the sum-rate maximization problem of interest is the following

%
\begin{align}
\label{P:sumRateMax1}
&\max \limits_{\mathcal{V},\mathcal{B}} R_{tot} \left(\mathcal{V},\mathcal{B}\right) = \max \limits_{\mathcal{V},\mathcal{B}} \sum\nolimits_{i = 1}^{N_{u}}\alpha_{i}R_{i}\left(\mathcal{V},\mathcal{B}\right) & \\
& \quad \text{s.t.} \quad \quad \tr \left(\mbf{V}_i\ \mbf{V}^H_i \right) \le P_{i}, \quad ~~ i=1,\dots,N_{u}  \tag{\ref{P:sumRateMax1}.a}\label{Pow_cons}\\
& \quad \Re \left({b}_{k,p}\right) = R_0 , \quad k=1,\dots,K, \; p=1,\dots,P \tag{\ref{P:sumRateMax1}.b}\label{IRS_cons1} \\
& \quad \Im\left({b}_{k,p}\right) \in \mathbb{R}, \quad k=1,\dots,K, \; p=1,\dots,P \tag{\ref{P:sumRateMax1}.c}\label{IRS_cons2} 
 \end{align}
\noindent where $R_0 \ge 0$ is a constant resistance that accounts for the losses of the tunable impedances of the RIS elements \cite{Gabriele_HE2E}.

The optimization problem in \eqref{P:sumRateMax1} is, however, not convex in the optimization variables $\mathcal{V}$ and $\mathcal{B}$. Thus, it is difficult to solve it globally. Then, we introduce a sub-optimal but tractable iterative algorithm to tackle it. The proposed approach is given in Algorithm \ref{Alg:DIA-PD} and is detailed in the next sub-sections. In general terms, at each iteration of Algorithm \ref{Alg:DIA-PD}, we first solve \eqref{P:sumRateMax1} as a function of $\mathcal{V}$ by assuming $\mathcal{B}$ fixed, and then we solve \eqref{P:sumRateMax1} as a function of $\mathcal{B}$ by assuming $\mathcal{V}$ fixed. Algorithm \ref{Alg:DIA-PD} combines, at each iteration, the solutions of the two sub-problems according to the BCD method.

\subsection{Precoding Optimization} \label{Precoding_opt}
We commence with the computation of $\mathcal{V}$ with $\mathcal{B}$ kept fixed. At the $q$th iteration of Algorithm \ref{Alg:DIA-PD}, this corresponds to executing Algorithm \ref{Alg:wMMSE} by setting  $\mathbf{B}_k = \mathbf{B}_k^{(q)}$. By assuming $\mathcal{B}$ fixed, the problem in \eqref{P:sumRateMax1} reduces to a conventional precoding optimization problem, which is, however, not jointly convex in the $N_u$ precoding matrices $\mathcal{V}$. To tackle it, we utilize the wMMSE algorithm \cite{Shi2011}, as summarized in Algorithm \ref{Alg:wMMSE}.\footnote{In Algorithm \ref{Alg:wMMSE}, $\mu_i$ denote the Lagrange multipliers of the optimization problem. They are chosen so that the power constraint in \eqref{Pow_cons} is fulfilled.}

\begin{algorithm}[!t]
\footnotesize
\caption{wMMSE for precoding optimization}
\textbf{Define:}	$\mbf{E}_{i}\left(\mathcal{V},\mbf{G}_{i},\mathcal{B}\right) = \mathbf{I}_L-2\Re\left(\mbf{G}_{i}^H\mbf{H}_{i,i}\left(\mathcal{B}\right)\mbf{V}_{i}\right)+ \sigma_i^2 \mbf{G}^H_{i}\mbf{G}_{i}
\hspace{2.75cm}+\sum\nolimits_{j = 1}^{N_{u}} \mbf{G}^H_{i}\mbf{H}_{i,j}\left(\mathcal{B}\right)\mbf{V}_{j}\mbf{V}_{j}^H\left[\mbf{H}_{i,j}\left(\mathcal{B}\right)\right]^H \mbf{G}_i$;

		\For {$i = 1,\ldots,N_u$}
		{
		$\mbf{J}_{i} =\sum\nolimits_{j = 1}^{N_{u}} \mbf{H}_{i,j}\left(\mathcal{B}\right)\mbf{V}_{j}^{(q)}\left(\mbf{V}_{j}^{(q)}\right)^H\left[\mbf{H}_{i,j}\left(\mathcal{B}\right)\right]^H  + \sigma_i^2 \mbf{I}_{L}$;\\
		$\mbf{G}_{i}^{(q+1)}=  \mbf{J}_{i}^{-1}\mbf{H}_{i,i}\left(\mathcal{B}\right)\mbf{V}_{i}^{(q)}$;\\
		$\mbf{W}_{i}^{(q+1)} =  \left[\mbf{E}_{i}\left(\mathcal{V}^{(q)},\mbf{G}^{(q+1)}_{i},\mathcal{B}\right)\right]^{-1}$;\\
		$\mbf{K} = \sum\nolimits_{j = 1}^{N_{u}}\alpha_{j} \left[\mbf{H}_{j,i}\left(\mathcal{B}\right)\right]^H \mbf{G}_{j}^{(q+1)} \mbf{W}_{j}^{(q+1)} \mbf{G}_{j}^{(q+1)} \mbf{H}_{j,i}\left(\mathcal{B}\right)$;\\
		$\mbf{V}_{i}^{(q+1)} =\alpha_{i}\left(\mbf{K}+ \mu_i \mbf{I}_{M} \right)^{-1}\left[\mbf{H}_{i,i}\left(\mathcal{B}\right)\right]^H\mbf{G}_{i}^{(q+1)}\mbf{W}_{i}^{(q+1)}$;
		}
		
\label{Alg:wMMSE} 
\end{algorithm} \setlength{\textfloatsep}{-15pt}%

\subsection{RIS Optimization -- Formulation and Challenges} \label{RIS_opt}
Subsequently, we solve the problem in \eqref{P:sumRateMax1} as a function of $\mathcal{B}$ by assuming $\mathcal{V}$ fixed. This corresponds to executing the inner loop (as a function of $K$) in Algorithm \ref{Alg:DIA-PD}. We utilize again the wMMSE algorithm at each iteration of Algorithm \ref{Alg:DIA-PD}. Unlike the application of the wMMSE algorithm in Section \ref{Precoding_opt}, the reformulation in terms of wMMSE framework is, however, not easy to solve in this case. To appreciate the difficulties of computing $\mathcal{B}$, let us first reformulate \eqref{P:sumRateMax1} according to the wMMSE framework. We obtain the following problem \cite{Shi2011}
\begin{align}
\label{wMMSE1_2}
&\min \limits_{\mathcal{B}} \sum\nolimits_{i = 1}^{N_{u}}\alpha_{i}\tr\left(\mbf{W}_{i}\mbf{E}_{i}\left(\mathcal{V},\mbf{G}_{i},\mathcal{B}\right)\right)&\\
& \text{s.t.} \quad 
\Re \left({b}_{k,p}\right) = R_0 , \quad k=1,\dots,K, \; p=1,\dots,P \tag{\ref{wMMSE1_2}.a} \label{Const_wmmSE_1} \\
& \, \, \, \, \, \, \, \, \, \, \, \, \, \,  \Im\left({b}_{k,p}\right) \in \mathbb{R}, \quad k=1,\dots,K, \; p=1,\dots,P \tag{\ref{wMMSE1_2}.b} \label{Const_wmmSE_2}
\end{align} 
\noindent where $\mbf{E}_{i}\left(\mathcal{V},\mbf{G}_{i},\mathcal{B}\right) = \mbf{E}_{i}^{(q)}\left(\mathcal{V}^{(q)},\mbf{G}_{i}^{(q)},\mathcal{B}^{(q)}\right)$ is defined in Algorithm \ref{Alg:wMMSE}, and $\mbf{G}_{i} = \mbf{G}_{i}^{(q)}$, $\mbf{W}_{i} = \mbf{W}_{i}^{(q)}$ and $\mbf{V}_{i} = \mbf{V}_{i}^{(q)}$ are the solutions of Algorithm \ref{Alg:wMMSE} at the $q$th iteration.

Compared with the wMMSE algorithm in Section \ref{Precoding_opt}, the main challenge for solving \eqref{wMMSE1_2} lies in the end-to-end matrices ${{{\bf{\tilde H}}}_{i,k,j}} $ in \eqref{HB} that depend on the inverse of the matrices of tunable impedances $\mathbf{B}_{k}$. This implies that $\mbf{E}_{i}\left(\cdot\right)$ is not convex in each of the optimization variables in $\mathcal{B}$ while keeping the others fixed. For completeness, we remark that the wMMSE algorithm has recently been utilized in \cite{Cunhua_wMMSE} and \cite{Andrea_wMMSE} for optimizing the sum-rate of RIS-assisted systems in the presence of instantaneous and statistical channel state information, respectively. In these latter papers, however, similar to Section \ref{Precoding_opt}, the corresponding $\mbf{E}_{i}\left(\cdot\right)$ matrices of the wMMSE algorithmic reformulation are convex in each of the optimization variables while keeping the others fixed. Therefore, the problem formulations in \cite{Cunhua_wMMSE} and \cite{Andrea_wMMSE} are easier to solve. In \cite{Cunhua_wMMSE} and \cite{Andrea_wMMSE}, in addition, the elements of the RISs are modeled as ideal unit-modulus phase shifters, and the impact of the mutual coupling and tunable circuits is not considered. These two aspects make the problem formulation in \eqref{wMMSE1_2} unique, and, to the best of our knowledge, the wMMSE reformulation in \eqref{wMMSE1_2} has never been tackled in the context of optimizing RIS-assisted MIMO interference channels. Finally, the constraints \eqref{Const_wmmSE_1} and \eqref{Const_wmmSE_2} are not the conventional unit-modulus constraints used in the literature, e.g., in \cite{Cunhua_wMMSE}, \cite{Andrea_wMMSE}.

\subsection{RIS Optimization -- Algorithmic Solution} \label{RIS_opt__Solution}

To solve the problem in \eqref{wMMSE1_2} by circumventing these issues, we leverage the Neuman series approximation \cite{Neuman_series_source}. Specifically, the inverse matrix in \eqref{HB} is calculated through a linearization, which allows us to tackle the non-convexity of $\mbf{E}_{i}\left(\cdot\right)$ with $\mathcal{B}$. In detail, at each iteration of Algorithm \ref{Alg:DIA-PD}, $\mathbf{B}^{(q)}_{k}$ is updated through small increments (perturbations). Let $\mathbf{\Delta}_k = \text{diag}\left(\boldsymbol{\delta}_{k}\right) \in \mathbb{C}^{P \times P}$ be the diagonal matrix of such a small perturbations for $k=1,2, \ldots, K$. The updating policy at each iteration is $\mathbf{B}^{(q+1)}_{k} = \mathbf{B}^{(q)}_{k} + \mathbf{\Delta}_k$. By defining $\mathbf{X}^{(q)}_{k} = \left(\bar{\mathbf{B}}_{k}+\mathbf{B}^{(q)}_{k}\right)^{-1}$, with the aid of the Neuman series approximation \cite{Neuman_series_source}, we obtain
\begin{equation}
\begin{array}{l}
\left(\bar{\mathbf{B}}_{k}+\mathbf{B}^{(q+1)}_{k} \right)^{-1} \approx \mathbf{X}^{(q)}_{k} - \mathbf{X}^{(q)}_{k} \mathbf{\Delta}_k \mathbf{X}^{(q)}_{k}
\end{array}
\label{eq:Neuman_approx}
\end{equation}
\noindent which is sufficiently accurate if $\left\|\mathbf{\Delta}_k \mathbf{X}^{(q)}_{k}  \right \| \ll 1$, where $\left\| \mathbf{X} \right \|$ is the spectral norm of $\mathbf{X}$, i.e., the largest eigenvalue of $\mathbf{X}^H\mathbf{X}$ \cite[Eq. (4.17)]{Neuman_series_source}. Since $\left\|\mathbf{\Delta}_k \mathbf{X}^{(q)}_{k}  \right \| \le \left\|\mathbf{\Delta}_k \right \| \left\|\mathbf{X}^{(q)}_{k}  \right \|$, the inequality $\left\|\mathbf{\Delta}_k \mathbf{X}^{(q)}_{k}  \right \| \ll 1$ is equivalent to $\left\|\mathbf{\Delta}_k\right \| = \delta/ \left\|\mathbf{X}^{(q)}_{k}  \right \|$ with $\delta \ll 1$. In Algorithm \ref{Alg:DIA-PD}, $\delta$ is set small enough to make \eqref{eq:Neuman_approx} accurate.

Thanks to the re-writing $\mathbf{B}^{(q+1)}_{k} = \mathbf{B}^{(q)}_{k} + \mathbf{\Delta}_k$, the problem in \eqref{wMMSE1_2} can be equivalently reformulated in terms of $\mathbf{\Delta}_k$ as optimization variables. In particular, the constraint in \eqref{Const_wmmSE_1} is enforced by setting $\Re( \mathbf{B}^{(0)}_{k}) = R_0 \mathbf{I}_P$ and considering only the imaginary part of $\mathbf{\Delta}_k$ for updating $\mathbf{B}^{(q)}_{k}$ at each iteration of Algorithm \ref{Alg:DIA-PD}. This is further elaborated and detailed next.

For ease of notation, let $\mathcal{D} = \left\{\boldsymbol{\delta}_{1},\ldots,\boldsymbol{\delta}_{K}\right\}$ denote the new set of optimization variables. Thanks to \eqref{eq:Neuman_approx}, the optimization problem in \eqref{wMMSE1_2} can be solved by applying again the wMMSE algorithm. By capitalizing on the linearization in \eqref{eq:Neuman_approx}, in particular, the optimization problem in \eqref{wMMSE1_2} is convex in the generic optimization variable $\boldsymbol{\delta}_{k}$ while keeping the other variables $\boldsymbol{\delta}_{i \ne k}$ fixed.  Therefore, the BCD-based method can be applied to obtain a locally optimal solution of \eqref{wMMSE1_2}. In particular, the impedance matrices of each RIS, $\mathbf{B}_{k}$, can be computed one-by-one in an iterative fashion as illustrated in the inner loop of Algorithm \ref{Alg:DIA-PD} and detailed next.

\subsection{RIS Optimization -- Closed-Form Formulation} \label{RIS_opt__SolutionB}
In order to compute $\mathbf{B}_{k}$ and solve the problem in \eqref{wMMSE1_2} with the aid of \eqref{eq:Neuman_approx}, we employ the wMMSE algorithm \cite{Shi2011}, whose specific implementation details are given in this section and correspond to the inner loop in Algorithm \ref{Alg:DIA-PD}. In particular, the objective of this section is to derive a closed-form analytical expression for the optimization variables $\mathcal{D}$. For ease of exposition and to leverage the BCD method, we introduce the notation $\mathcal{D}_{\sim k} = \left\{\boldsymbol{\delta}_{1},\ldots,\boldsymbol{\delta}_{k-1},\boldsymbol{\delta}_{k+1},\ldots,\boldsymbol{\delta}_{K}\right\}$ that yields the set of all optimization variables with the exception of $\boldsymbol{\delta}_k$. Based on this notation, $\mbf{E}_{i}\left(\mathcal{V},\mbf{G}_{i},\mathcal{B}\right)$ can be formulated as
\begin{equation}
\begin{array}{l}
\mbf{E}_{i}\left(\mathcal{D}\right) = \mbf{E}_{i,k}\left(\boldsymbol{\delta}_k,\mathcal{D}_{\sim k}\right) + \Upsilon\left(\mathcal{D}_{\sim k}\right)
\end{array}
\label{eq:eq_Eik1}
\end{equation}
where $\Upsilon\left(\mathcal{D}_{\sim k}\right)$ collects all terms that are independent of $\boldsymbol{\delta}_k$.

Based on the BCD-based method \cite{Shi2011}, at each iteration of Algorithm \ref{Alg:DIA-PD}, the variable $\boldsymbol{\delta}_k$ is computed by assuming that $\mathcal{D}_{\sim k}$ is fixed. At the $k$th iteration of the inner loop in Algorithm \ref{Alg:DIA-PD}, we are interested in $\mbf{E}_{i,k}\left(\boldsymbol{\delta}_k,\mathcal{D}_{\sim k}\right)$, while $\Upsilon\left(\mathcal{D}_{\sim k}\right)$ can be disregarded. After lengthy algebraic manipulations that are omitted due to space limitations and ignoring some constant terms that are irrelevant to the optimization problem, at the $q$th iteration of Algorithm \ref{Alg:DIA-PD}, $\mbf{E}_{i,k}\left(\boldsymbol{\delta}_k,\mathcal{D}_{\sim k}\right)$ can be formulated as
\begin{align}
& \mbf{E}_{i,k}^{(q)} \left(\boldsymbol{\delta}_{k},\mathcal{D}_{\sim k}\right) = \sum\nolimits_{j=1}^{N_u }  \left[\mathbf{A}_{i,k}^{(q)} \mathbf{\Delta}_{k} \mathbf{C}_{k,j}^{(q)} \right]^+ -2 \Re \left ( \mathbf{A}_{i,k}^{(q)} \mathbf{\Delta}_{k} \mathbf{C}_{k,i}^{(q)} \right ) \nonumber \\
& \hspace{-0.18cm}+ 2 \Re \left ( \sum\nolimits_{j=1}^{N_u }\mathbf{A}_{i,k}^{(q)}  \mathbf{\Delta}_{k} \mathbf{D}_{i,k,j}^{(q)} \right )  +
2 \Re \left ( \sum\nolimits_{j=1}^{N_u } \mathbf{A}_{i,k}^{(q)} \mathbf{\Delta}_{k}\mathbf{F}_{i,k,j}^{(q)} \right ) 
\label{eq_MMSE_new}
\end{align}
\noindent where $\mathbf{A}_{i,k}^{(q)} = -\mbf{G}^H_{i}  \mathbf{T}_{i,k} \mathbf{X}^{(q)}_{k} \in \mathbb{C}^{L \times P}$, $\mathbf{C}_{k,j}^{(q)} = \mathbf{X}^{(q)}_{k}\mathbf{S}_{k,j} \mathbf{V}_{j} \in \mathbb{C}^{P \times L}$, 
$\mathbf{D}_{i,k,j}^{(q)} = \mathbf{X}^{(q)}_{k}\mathbf{S}_{k,j} \mathbf{V}_{j}^+ \left(\mathbf{{\hat{H}}}_{i,j}^{(q)}\right)^H\mbf{G}_{i} \in \mathbb{C}^{P \times L}$, $\mathbf{X}^+ = \mathbf{X}\mathbf{X}^H$,
\begin{equation}
    \mathbf{{\hat{H}}}_{i,j}^{(q)} = \mathbf{{\bar{H}}}_{i,j} - \sum\nolimits_{k=1}^{K}\mathbf{T}_{i,k}\mathbf{X}^{(q)}_{k} \mathbf{S}_{k,j} \quad \in \mathbb{C}^{L \times L} 
\end{equation}
\begin{equation}
   \mathbf{F}_{i,k,j}^{(q)} = \mathbf{X}^{(q)}_{k} \mathbf{S}_{k,j}\mathbf{V}_{j}\left(\mbf{G}^H_{i}\sum\nolimits_{m=1 \atop m \ne k}^{K} \mathbf{\tilde{H}}_{i,m,j} \left(\mbf{b}_{m}\right)\mathbf{V}_j\right)^H  \in \mathbb{C}^{P \times L} .
\end{equation}

By direct inspection of \eqref{eq_MMSE_new}, we evince that $\mbf{E}_{i,k}^{(q)} \left(\boldsymbol{\delta}_{k},\mathcal{D}_{\sim k}\right)$ is a convex function in $\boldsymbol{\delta}_{k}$ if $\mathcal{D}_{\sim k}$ if kept fixed. The locally optimal solution of $\boldsymbol{\delta}_{k}$ we are looking for can, therefore, be obtained by minimizing the objective function in \eqref{wMMSE1_2}. To this end, the gradient of $\mbf{E}_{i,k}^{(q)} \left(\boldsymbol{\delta}_{k},\mathcal{D}_{\sim k}\right)$ is needed. In particular, the gradient of $\mbf{E}_{i,k}^{(q)} \left(\boldsymbol{\delta}_{k},\mathcal{D}_{\sim k}\right)$ needs to be computed with respect to the imaginary part of $\boldsymbol{\delta}_{k}$, i.e., $\boldsymbol{\delta}^{(I)}_k = \Im\left({\boldsymbol{\delta}_k}\right)$, in order to fulfill the constraint in \eqref{Const_wmmSE_1}, as discussed in previous text.

In order to compute the gradient of $\mbf{E}_{i,k}^{(q)} \left(\boldsymbol{\delta}_{k},\mathcal{D}_{\sim k}\right)$ in \eqref{eq_MMSE_new} with respect to $\boldsymbol{\delta}^{(I)}_k$, we introduce the compact indexing notation $\mathbf{M}^{\left\{n\text{:}k,l\text{:}p\right\}}$, which yields the submatrix extracted from the $n$th to the $k$th rows and from the $l$th to the $p$th columns of $\mathbf{M}$. Also, we introduce the mapping $\mathbf{Q}\left(\mathbf{Q}_1,\mathbf{Q}_2\right)$ between the matrices $\mathbf{Q}_1 \in \mathbb{C}^{L \times P}$ and $\mathbf{Q}_2 \in \mathbb{C}^{P \times L}$ and the matrices $\mathbf{Q} \in \mathbb{C}^{L \times PL}$ and $\boldsymbol{\Gamma}_k \in \mathbb{C}^{PL \times L}$ defined as
\vspace{-0.2cm}
\begin{equation}
\begin{array}{ll}
\mathbf{Q}^{\left\{l,(k-1)P+1\text{:}kP\right\}} (\mathbf{Q}_1,\mathbf{Q}_2)= \mathbf{Q}_1^{\left\{l,\text{:}\right\}}   \odot \left(\mathbf{Q}_2^{\left\{\text{:},k\right\}}\right)^T \\
\boldsymbol{\Gamma}_k ^{\left\{(l-1)P+1\text{:}lP,l\right\}} = \boldsymbol{\delta}_k 
\end{array}
\label{eq:rx_rearrange} \vspace{-0.1cm}
\end{equation}
\noindent which fulfill the identity $\mathbf{Q}_1 \mathbf{\Delta}_{k} \mathbf{Q}_2 = \mathbf{Q}\left(\mathbf{Q}_1,\mathbf{Q}_2\right) \boldsymbol{\Gamma}_k$.

Based on \eqref{eq:rx_rearrange}, $\mbf{E}_{i,k}^{(q)} \left(\boldsymbol{\delta}_{k},\mathcal{D}_{\sim k}\right)$ in \eqref{eq_MMSE_new} can be re-written as
\vspace{-0.2cm}
\begin{align}
&\mbf{E}_{i,k}^{(q)} \left(\boldsymbol{\delta}_{k},\mathcal{D}_{\sim k}\right) = \sum\nolimits_{j=1}^{N_u } \left[\mathbf{{Q}}\left(\mathbf{A}_{i,k}^{(q)},\mathbf{C}_{k,j}^{(q)}\right)\boldsymbol{\Gamma}_{k}\right]^+ \nonumber \\
& \hspace{1.25cm} -2 \Re \left(\mathbf{{Q}}\left(\mathbf{A}_{i,k}^{(q)},\mathbf{C}_{k,i}^{(q)}\right)\boldsymbol{\Gamma}_{k}\right ) \\
& \hspace{1.25cm} + 2 \Re \left(\sum\nolimits_{j=1}^{N_u }\mathbf{{Q}}\left(\mathbf{A}_{i,k}^{(q)},\mathbf{D}_{i,k,j}^{(q)}\right)+\mathbf{{Q}}\left(\mathbf{A}_{i,k}^{(q)},\mathbf{F}_{k,j}^{(q)}\right)\right)\boldsymbol{\Gamma}_{k} \nonumber .
\label{eq:mMSEMat4.11} \vspace{-0.2cm}
\end{align}

For mathematical convenience, we introduce the notation
\begin{align}
& \mathbf{\tilde{M}}_{k}^{(q)} = \sum\nolimits_{i=1}^{N_u} \alpha_i \sum\nolimits_{j=1}^{N_u }  \mathbf{{Q}}^H\left(\mathbf{A}_{i,k}^{(q)},\mathbf{C}_{k,j}^{(q)}\right) \mathbf{W}_i \mathbf{{Q}}\left(\mathbf{A}_{i,k}^{(q)},\mathbf{C}_{k,j}^{(q)}\right) \nonumber \\
& \left(\mathbf{\tilde{U}}_{k}^{(q)}\right)^H  \hspace{-0.2cm}=\hspace{-0.1cm} \sum\nolimits_{i=1}^{N_u} \alpha_i  \mathbf{W}_i \sum\nolimits_{j=1}^{N_u } \mathbf{{Q}}\left(\mathbf{A}_{i,k}^{(q)},\mathbf{C}_{k,j}^{(q)}\right)-\mathbf{{Q}}\left(\mathbf{A}_{i,k}^{(q)},\mathbf{D}_{i,k,j}^{(q)}\right) \nonumber \\ 
& \hspace{1.1cm} - \mathbf{{Q}}\left(\mathbf{A}_{i,k}^{(q)},\mathbf{F}_{i,k,j}^{(q)}\right)  \, .
\vspace{-0.2cm}
\end{align}

Based on these definitions, the gradient of the objective function in \eqref{wMMSE1_2} (at the $q$th iteration of Algorithm \ref{Alg:DIA-PD}) with respect to $\boldsymbol{\delta}^{(I)}_k$ can be formulated as follows
\vspace{-0.2cm}
\begin{equation}
\label{wMMSE2r}
\begin{aligned}
& \nabla_{\boldsymbol{\delta}^{(I)}_k} \sum\nolimits_{i = 1}^{N_{u}}\alpha_{i}  \tr\left(\mbf{W}_{i}^{(q)}\mbf{E}_{i}^{(q)}\left(\boldsymbol{\delta}_{k},\mathcal{D}_{\sim k}\right)\right) 
\\ & \hspace{3.5cm} = 2\Im \left(\mathbf{{M}}_{k}^{(q)}\right) \boldsymbol{\delta}^{(I)}_k  - 2 \Im\left(\mathbf{{u}}_{k}^{(q)}\right)
\end{aligned} \vspace{-0.2cm}
\end{equation}
\noindent where $\mathbf{{M}}_{k}^{(q)} \in \mathbb{C}^{P \times P} $ and $\mathbf{{u}}_{k}^{(q)}  \in \mathbb{C}^{P \times 1} $ are defined as
\begin{equation}
\begin{array}{ll}
& \mathbf{{M}}_{k}^{(q)} = \sum\nolimits_{l=1}^{L} \left(\mathbf{\tilde{M}}_{k}^{(q)}\right) ^{\left\{(l-1)P+1\text{:}lP,(l-1)P+1\text{:}lP\right\}}\\
& \mathbf{{u}}_{k}^{(q)} = \sum\nolimits_{l=1}^{L} \left(\mathbf{\tilde{U}}_{k}^{(q)}\right)^{\left\{(l-1)P+1\text{:}lP,l\right\}}
\end{array}
\label{eq:mMSEMat4.4} \, . \vspace{-0.1cm}
\end{equation}

From \eqref{wMMSE2r}, we obtain the solution
\begin{equation}
\label{eq:precoderShir}
\boldsymbol{\delta}_{k}^{(I)} = \left(\Im \left(\mathbf{{M}}_{k}^{(q)}\right) + \mu_k \right)^{-1}{\Im \left(\mathbf{{u}}_{k}^{(q)}\right)}
\end{equation}
where the Lagrange multipliers $\mu_k$ are chosen to fulfill the constraint $\left\|\mathbf{\Delta}_k\right \| = \delta/ \left\|\mathbf{X}^{(q)}_{k}  \right \|$. From \eqref{eq:precoderShir}, the matrices of tunable impedances are iteratively updated as $\mathbf{B}^{(q+1)}_{k}  = \mathbf{B}^{(q)}_{k} + \mathbf{\Delta}_k$.

\subsection{Convergence of Algorithm \ref{Alg:DIA-PD}}
Algorithm \ref{Alg:DIA-PD} consists of the iterative solution of a set of convex subproblems that are characterized by a global utility function, i.e., the weighted mean square error (wMSE). This ensures that the wMSE is a non-increasing function of $q$. Since the wMSE is lower bounded by zero, Algorithm \ref{Alg:DIA-PD} is monotonically convergent. The convergence of the weighted sum-rate $R_{tot} \left(\mathcal{V},\mathcal{B}\right)$ in \eqref{P:sumRateMax1} is analyzed as follows.
\vspace{-0.25cm}
\begin{prop}\label{conv_proof}
$R_{tot} \left(\mathcal{V},\mathcal{B}\right)$ is a non-decreasing function in $q$. 
\end{prop} \vspace{-0.19cm}
\begin{IEEEproof}
Denote $\hat{\mbf{E}}_{i}^{(q)} = \mbf{E}_{i}\left(\mathcal{V}^{(q)},\mbf{G}^{(q)}_i,\mathcal{B}^{(q)}\right)$ at the $q$th iteration of Algorithm \ref{Alg:DIA-PD}. From \cite{Shi2011}, the rate $R^{(q)}_i$ in \eqref{P:sumRateMax1} can be expressed as $R^{(q)}_i = \log \det\left(\hat{\mbf{E}}_{i}^{(q)}\right)^{-1}$ and the optimal weights in Algorithm \ref{Alg:wMMSE} can be expressed as $\mbf{W}^{(q)}_{i} = \left(\hat{\mbf{E}}_{i}^{(q)}\right)^{-1}$. The wMSE at the $q$th iteration is, therefore, equal to $LN_u-R^{(q)}_i$, and $R_{tot} \left(\mathcal{V},\mathcal{B}\right)$ is a non-decreasing function of $q$. 
\end{IEEEproof}

\begin{figure*}\hspace{-1cm}
\begin{minipage}[t]{0.34\linewidth}
\centering
\includegraphics[scale=.45]{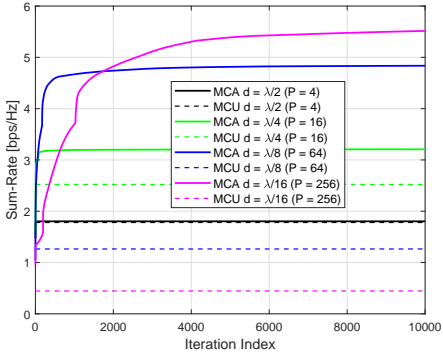}\\
\caption{Convergence of Algo. \ref{Alg:DIA-PD} ($L = 5$).}
\label{fig1}
\end{minipage}
\begin{minipage}[t]{0.02\linewidth}
\end{minipage}
\begin{minipage}[t]{0.33\linewidth}
\centering
\includegraphics[scale=.45]{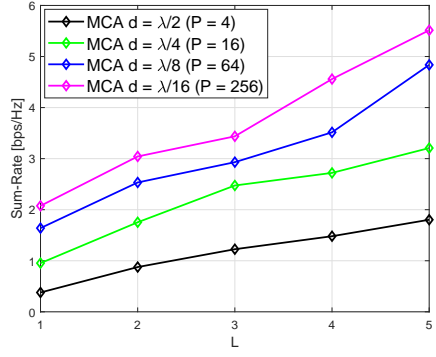}\\
\caption{Sum-rate vs. $L$ (10000 iterations).}
\label{fig2}
\end{minipage}
\begin{minipage}[t]{0.02\linewidth}
\end{minipage}
\begin{minipage}[t]{0.33\linewidth}
\centering
\includegraphics[scale=.45]{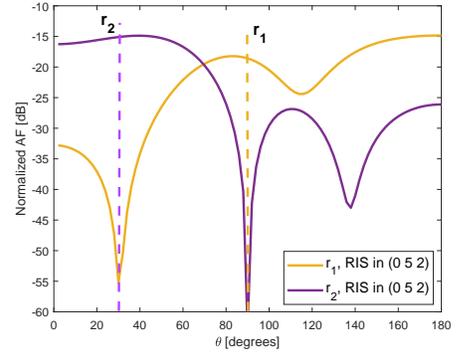}\\
\caption{Equivalent AF ($d=\lambda/16$).}
\label{fig3}
\end{minipage}
\vspace{-0.3cm}
\end{figure*}

\section{Numerical Results}
\label{sec:NumericalResults}
We consider a setup with two transmitter-receiver pairs ($N_u =  2$) located in $\mathbf{t}_1 = \left(5~20~1\right)$, $\mathbf{t}_2 = \left(5~10~1\right)$,
$\mathbf{r}_1 = \left(5~5~1\right)$ and
$\mathbf{r}_2 = \left(5~25~1\right)$ and two RISs ($K = 2$) centered in $\left(0~20~2\right)$ and $\left(0~5~2\right)$. The transmission frequency is $f = 28$ GHz and the wavelength is $\lambda$. The number of antennas at the transmitters and receivers is the same, i.e., $L = M$, and their inter-distance is $\lambda/2$. The RIS scattering elements are thin wires with radius $a = \lambda/500$ and length $l = \lambda/32$. Also, $R_0 = 0.2$ Ohm. To assess the impact of having sub-wavelength inter-distances while keeping the simulation time reasonably short, we assume that the size of each RIS is fixed to $\lambda \times \lambda$, which may represent a super-cell in a large-size RIS. Thus, the number of scattering elements $P$ and their inter-distances are chosen appropriately, e.g., $P = \left\{4,16,64,256\right\}$ for $d = \left\{1/2,1/4,1/8,1/16\right\}\lambda$. The noise power and transmit power are $\sigma^{2}_{i} = -120\,$dBm and $P_i = 20\,$dBm. The matrices of mutual coupling are computed as detailed in \cite[Lemma 2]{Gabriele_HE2E}. The transmitters and receivers are assumed to be in non-line-of-sight so that the direct links are ignored. For comparison, two case studies are considered: (i) the \emph{mutual coupling aware} (MCA) setup in which the mutual coupling is taken into account at the optimization stage; and (ii) the \emph{mutual coupling unaware} (MCU) setup in which the off-diagonal entries of $\bar{\mathbf{B}}_{k}$ are set equal to zero at the optimization stage but are considered when computing the resulting sum-rate (i.e., mismatched design).

In Fig. \ref{fig1}, we observe the convergence of Algorithm \ref{Alg:DIA-PD} according to Proposition \ref{conv_proof}. We note the important role played by presence of mutual coupling in sub-wavelength designs and the need of taking it into account at the optimization stage. 
For values of inter-distances up to $d=\lambda/4$, increasing the number of scattering elements may compensate in part for the negative impact of mutual coupling, but this does not hold for smaller inter-distances. In Fig. \ref{fig2}, we observe that, if mutual coupling is taken into account, increasing the number of antennas at the transmitters and receivers enhances the sum-rate, and RISs with closely-spaced scattering elements yield superior performance. 
In Fig. \ref{fig3}, we report the equivalent array factor (AF) of the RIS centered in $\left(0~5~2\right)$ assuming $d=\lambda/16$. The AF is obtained, as a function of the angle of view ($\theta$) of the considered RIS, after configuring the RIS by using Algorithm \ref{Alg:DIA-PD}. Based on the considered network topology, we evince that Algorithm \ref{Alg:DIA-PD} configures the RIS so that its AF has a null in correspondence of the angle under which the RIS views the interfered receiver for each considered intended link.

\vspace{-0.03cm}
\section{Conclusion} \vspace{-0.1cm}
We have introduced a provably convergent optimization algorithm for maximizing the sum-rate of RIS-assisted MIMO interference channels. The proposed approach accounts for the mutual coupling among closely-spaced scattering elements. Numerical results have validated the convergence of the proposed approach and the need of accounting for the mutual coupling among the scatterers of the RIS at the design stage.

\vspace{-0.03cm}
\bibliographystyle{IEEEtran}

\end{document}